\documentclass[12pt]{article}
\usepackage{fullpage}
\usepackage[utf8]{inputenc}
\usepackage{amsmath, biometri, mathtools}
\usepackage{amssymb}
\usepackage{amsfonts,}
\usepackage{bbm}
\usepackage{graphicx}
\usepackage{float}
\usepackage{caption}
\usepackage{subcaption,mathrsfs,multirow}
\usepackage[ruled]{algorithm2e}
\usepackage{array}
\def\Dbb{\mathbb{D}}
\usepackage{xr}
\newtheorem{remark}{Remark}

\usepackage{verbatim}
\usepackage{listings}
\usepackage{color}
\usepackage{epsfig}
\usepackage{wasysym}

\usepackage{color}

\newcommand{\comout}[1]{}
\def\Sbb{\mathbb{S}}

\def\Dscr{\mathscr{D}}

\def\subopt{_{\mbox{\tiny opt}}}

\def\gopt{g\subopt}
\def\gtildeopt{\tilde{g}\subopt}
\def\subnew{_{\mbox{\tiny new}}}
\def\Psc{\mathcal{P}}

\def\Lsc{\mathcal{L}}

\def\lpbox#1{\vskip1mm \begin{center}
        \hspace{.0\textwidth}\vbox{\hrule\hbox{\vrule\kern6pt
\parbox{.95\textwidth}{\kern6pt \blue #1 (LP)\vskip6pt}\kern6pt\vrule}\hrule}
        \end{center} \vskip-5mm}

\usepackage[usenames,dvipsnames,svgnames,table]{xcolor}
\usepackage[colorlinks=true,
            linkcolor=red,
            urlcolor=blue,
            citecolor=blue]{hyperref}

\graphicspath{{./}{./Figures/}}

\usepackage{setspace,tcolorbox}
\definecolor{lbcolor}{rgb}{0.95,0.95,0.95}
\def\Fhat{\widehat{F}}
\def\Fscr{\mathscr{F}}

\def\EE{\mathbb{E}}

\def\half{\frac{1}{2}}

\def\Fbb{\mathbb{F}}
\def\Gbb{\mathbb{G}}
\def\Lscr{\mathscr{L}}
\def\Isc{\mathcal{I}}
\def\subIsc{_{\scriptscriptstyle \Isc}}

\def\Mbb{\mathbb{M}}

\def\PTE{\mbox{PTE}}

\def\RE{\mbox{RE}}
\def\PTEhat{\widehat{\mbox{PTE}}}

\def\lambdahat{\widehat{\lambda}}
\def\ghat{\widehat{g}}
\def\mhat{\widehat{m}}
\def\fhat{\widehat{f}}

\def\muhat{\widehat{\mu}}
\def\sumin{\sum_{i=1}^n}

\def\Deltahat{\widehat{\Delta}}
\def\nhalf{n^{1/2}}
\def\nnhalf{n^{-1/2}}

\def\supone{^{(1)}}
\def\supzero{^{(0)}}
\def\supa{^{(a)}}
\def\Lsc{\mathcal{L}}

\def\ninv{n^{-1}}
\def\nhalf{n^{\half}}
\def\nnhalf{n^{-\half}}

\def\bD{\mathbf{D}}

\definecolor{darkred}{RGB}{150,50,50}
\definecolor{brown}{RGB}{250,100,100}
\definecolor{green}{RGB}{000,150,100}
\definecolor{purple}{RGB}{200,000,250}

\def\blue{\color{blue}}

\def\blue{\color{blue}}

\def\trans{^{\scriptscriptstyle \sf T}}

\def\subgopt{_{\gopt}}
\def\RE{\mbox{RE}}

\def\RP{\mbox{RP}}
\def\RPhat{\widehat{\RP}}

\def\subRPg{_{\scriptscriptstyle \sf RP_g}}
\def\subRPgopt{_{\scriptscriptstyle \sf RP_{\gopt}}}

\def\subRPgopti{_{\scriptscriptstyle {\sf RP}_{\gopt},i}}
\def\sigmahat{\widehat{\sigma}}
\def\bSigma{\boldsymbol{\Sigma}}
\def\psihat{\widehat{\psi}}
\def\subIsck{_{\scriptscriptstyle \Isc_k}}
\def\subIscnk{_{\scriptscriptstyle \Isc_{\text{-}k}}}

\def\supnk{^{\scriptscriptstyle (\text{-}k)}}
\def\subcv{_{\scriptscriptstyle \sf CV}}
\def\subghat{_{\ghat}}

\begin{document}

\begin{center}
{\large \bf Towards Optimal Use of Surrogate Markers to Improve Power} \vspace{.1in}

Xuan Wang \\
{\em Department of Biostatistics, Harvard University,  Boston, MA 02115 U.S.A.} \\[2ex]

Layla Parast \\
{\em Department of Statistics and Data Sciences, University of Texas at Austin, Austin, TX 78712 U.S.A.}\\[2ex]

{Lu Tian} \\
{\em Department of Biomedical Data Science, Stanford University, Stanford, CA 94305 U.S.A.} \\[2ex]

Tianxi Cai \\
{\em Department of Biostatistics and Department of Biomedical Informatics, Harvard University,  Boston, MA 02115 U.S.A. }

\end{center}

\begin{abstract}
Motivated by increasing pressure for decision makers to shorten the time required to evaluate the efficacy of a treatment such that treatments deemed safe and effective can be made publicly available, there has been substantial recent interest in using an earlier or easier to measure surrogate marker, $S$, in place of the primary outcome, $Y$. To validate the utility of a surrogate marker in these settings, a commonly advocated measure is the proportion of treatment effect on the primary outcome that is explained by the treatment effect on the surrogate marker (PTE). Model based and model free estimators for PTE have also been developed. While this measure is very intuitive, it does not directly address the important questions of how $S$ can be used to make inference of the unavailable $Y$ in the next phase clinical trials. In this paper, to optimally use the information of surrogate S, we provide a framework for deriving an optimal transformation of $S$, $\gopt(S)$, such that the treatment effect on $\gopt(S)$ maximally approximates the treatment effect on $Y$ in a certain sense. Based on the optimally transformed surrogate, $\gopt(S)$, we propose a new measure to quantify surrogacy, the relative power (RP), and demonstrate how RP can be used to make decisions with $S$ instead of $Y$ for next phase trials. We propose nonparametric estimation procedures, derive asymptotic properties, and compare the RP measure with the PTE measure. Finite sample performance of our estimators is assessed via a simulation study.  We illustrate our proposed procedures using an application to the Diabetes Prevention Program (DPP) clinical trial to evaluate the utility of hemoglobin A1c and fasting plasma glucose as surrogate markers for diabetes.

\vspace{10pt}

\end{abstract}

\newpage
\clearpage

\section{Introduction}

Motivated by increasing pressure for decision makers to shorten the time required to evaluate the efficacy of a treatment such that treatments deemed safe and effective can be made publicly available, there has been substantial recent interest in using an earlier or easier to measure surrogate marker in place of a primary outcome. The development and testing of clinical treatments, including vaccines, often require years of research and participants follow-up. Though strict and regulated testing is essential to guarantee that treatments are safe and effective, early indications about the effectiveness of the treatment based on a surrogate marker could potentially be used to make inference about the treatment effect on the primary outcome. The use of a surrogate marker in this way may allow for early testing of a treatment effect and lead to reduced follow up time and/or costs. For example, during the COVID-19 public health emergency in 2020, the Food and Drug Administration issued guidance allowing for an emergency use authorization for vaccines demonstrating efficacy with respect to a surrogate marker that is ``reasonably likely to predict" protection against COVID-19 (\cite*{avorn2020up,FDA20}). These urgent needs highlight the importance of developing methods to identify valid surrogate markers such that they may be used in future studies.

 For decades, the statistical, epidemiological, and clinical research communities have made substantial progress by proposing and evaluating methods to assess the value of potential surrogate markers (\cite*{prentice1989surrogate,molenberghs2002statistical,alonso2004validation,burzykowski2005evaluation,frangakis2002principal,gilbert2008evaluating,huang2011comparing,vanderweele2013surrogate,price2018estimation}). A formal definition for a valid surrogate marker was proposed in \cite{prentice1989surrogate} and since then, numerous methods have been proposed to validate surrogate markers or quantify the surrogacy of such surrogate markers. For example, \cite{freedman1992statistical} proposed a measure for the proportion of treatment effect on the primary outcome that is explained by the treatment effect on the surrogate (PTE) by examining the change in the treatment coefficient in a regression model predicting the primary outcome from the treatment with vs. without the surrogate marker included in the model. As a more flexible alternative, \cite{wang2002measure} proposed to quantify the PTE by evaluating what the treatment would be if the surrogate marker in the treatment group had the same distribution as the surrogate in the control group. While useful, these methods are model based and  lead to biased estimates of the PTE under model misspecification. A robust nonparametric model free estimation method was proposed by \cite{parast2016robust} to estimate the PTE defined by \cite{wang2002measure}. However, this method requires a monotone relationship between the outcome and the surrogate marker. Recently, \cite{wang2020model} proposed a model free strategy to quantify PTE that involves identifying an optimal transformation of the surrogate marker that best predicts the treatment effect on the primary outcome. This method is very robust and provides a way to infer the treatment effect on the primary outcome by using the optimally transformed surrogate marker. The derivation of the optimal transformation function relies on an working independence assumption, though the forms of the optimal transformation and PTE are not sensitive to the departure of the assumption. Note that these
quantities were proposed for a single study setting, different from a meta analytic setting where multiple studies
are available to investigate the surrogate marker and alternative measures have been developed to
validate surrogacy (\cite*{daniels1997meta,buyse1998criteria,burzykowski2005evaluation}).

In this paper, to avoid the working independence assumption that may be questioned, we derive an optimal transformation of the surrogate, $\gopt(\cdot)$, such that the treatment effect on $\gopt(S)$ maximally approximates the treatment effect on the primary outcome from another different aspect but as a supplementary to the optimal transformation in \cite{wang2020model}. The form of $\gopt(S)$ is analogous to the optimal transformation derived in \cite{wang2020model}. Simulation studies show that these two optimal transformations performs similarly. This also justify the robustness of the optimal transformation function in \cite{wang2020model}, which is relatively easier to implement. For simplicity we use notation $\gopt(S)$ for the optimal transformation in this paper, same as that in \cite{wang2020model}. 

The PTE quantity based on the proposed optimal transformation of the surrogate can provide useful information regarding the strength of a potential surrogate within, for example, a Phase 2 clinical trial, where testing is often conducted in a small number of patients in order to assess safety, monitor how a drug is metabolized, and gather initial data on efficacy. 
In the next clinical trial, such as Phase 3 clinical trial which is a large trial in patients to test efficacy and safety that provide the key data on efficacy in submissions for regulatory approval, one may be interested in understanding how this surrogate marker can be used to to make inference about the treatment effect on the primary outcome. That is, knowing that a particular surrogate marker explains, for example, 90\% of the treatment effect in an existing trial (Phase 2), what can be expected in a future trial (Phase 3) with respect to effect size and power, if that surrogate is used to make inference about the treatment effect instead of the primary outcome? With respect to using a surrogate marker to test for a treatment effect, useful methods have been proposed to improve power through the use of the surrogate marker information, when \textit{combined} with the primary outcome (\cite*{pepe1992inference,robins1992recovery,rotnitzky1995semiparametric,venkatraman1999properties,parast2014landmark}). Some recent work has addressed the question of how one can use a surrogate marker to \textit{replace} a primary outcome in a future study. For example, in a setting with multiple surrogate markers, \cite{athey2019surrogate} proposed a model-based approach to combining surrogate markers into a surrogate index that can be used to predict a treatment effect on the primary outcome. In a survival setting, \cite{parast2019using} proposed a testing procedure to test for a treatment effect using a single surrogate marker measured earlier in time. Importantly, this testing procedure requires a similar monotonicity assumption as  \cite{parast2016robust}, discussed earlier.

In this paper, we propose the relative power (RP) measure as an alternative measure of surrogacy, and discuss how this measure compared to existing approaches/measures. Our proposed measure aims to quantify the feasibility of using surrogate marker information to make inference about the primary outcome in a subsequent study. In addition, we demonstrate how this quantity can be used to inform future trial design. 

We propose robust nonparametric estimation procedures for $\gopt(\cdot)$, PTE and the RP measures and derive asymptotic properties of our estimators. Simulation results suggest that the proposed estimators perform well in finite sample.  We illustrate our approach using an application to the Diabetes Prevention Program (DPP) study where we examine two potential surrogate markers for diabetes, hemoglobin A1c and fasting plasma glucose.

\def\Fbb{\mathbb{F}}
\def\Gbb{\mathbb{G}}

\section{Identifying and Estimating an Optimal Transformation  \label{optimal}}

\subsection{Notation, Setting, and Assumptions \label{notation}}

Let $Y$ denote the primary outcome and $S$ be the surrogate marker such that $S$ can either be measured earlier than $Y$ or at the same time as $Y$ but with less cost or patient burden. The surrogate marker $S$ can be discrete or continuous; we treat $S$ as continuous for conciseness of presentation but the proposed methods can be easily modified to accommodate discrete $S$. Under the standard causal inference framework, let $Y^{(a)}$ and $S^{(a)}$ denote the respective potential outcome and potential surrogate under treatment $A = a \in \{0,1\}$. In practice, $(Y^{(1)}, S^{(1)})$ and $(Y^{(0)}, S^{(0)})$ cannot both be observed for  the same subject. We assume that  treatment assignment is random and without loss of generality $P(A = a) =0.5$. The observable data for analysis consist of $n$ sets of independent and identically distributed random vectors $\Dscr = \{\bD_i = (Y_i, S_i, A_i)\trans, i = 1, ..., n\}$, where  $Y_i = Y_i^{(1)}A_i + Y_i^{(0)}(1-A_i)$, $S_i = S_i^{(1)}A_i + S_i^{(0)}(1-A_i)$ and $n$ is the sample size.
The treatment effect on the primary outcome, $\Delta$ is defined as:
$$\Delta= \mu_1 - \mu_0 , \quad \mbox{where}\quad \mu_a= \EE(Y^{(a)}).$$
Without loss of generality, we assume $\Delta\geq 0$, which can always be realized by switching the two different treatment groups for analytic purposes if needed.

\def\Deltazeroone{\Delta_{01}}
\def\Deltahatzeroone{\Deltahat_{01}}

\subsection{Identifying $\gopt$}
It is desirable to identify an optimal prediction function such that the resulting $g(s)$ maximally predicts $Y$ while ensuring that $\Delta_g \le \Delta$ to maintain a desirable interpretation of $\Delta_g$, which is the treatment effect on $g(S)$, $\Delta_{g} = \mu_{g,1}  - \mu_{g,0}= \EE\{g(S^{(1)})-\EE\{g(S^{(0)})\}. $\cite{wang2020model}  identified an optimal $g$ that minimizes the mean squared error:
\begin{eqnarray*}
\Lscr_{\mbox{\tiny oracle}}(g) = \EE\left[(Y\supone - Y\supzero)-\{ g(S\supone)- g(S\supzero)\}\right]^2
\end{eqnarray*}
under the working independence assumption $ (Y^{(1)}, S^{(1)}) \perp  (Y^{(0)}, S^{(0)})$. This assumption is needed because the correlation between $(Y\supone, S\supone)$ and $(Y\supzero, S\supzero)$ is not identifiable. Although the inference procedures proposed in \cite{wang2020model} for quantifying the PTE of $g(S)$
do not require this  assumption to hold and the form of the optimal transformation is not sensitive to the departure of the assumption, the optimality of the resulting transformation may not hold when the working independence assumption is violated.

To overcome this challenge, we propose in this paper an alternative optimal $g$ that does $\textit{not}$ rely on this assumption. 
To this end, we note that maximizing $\PTE_g = \Delta_g/\Delta$ under the constraint of $\PTE_g \le 1$ is equivalent to minimizing
$\epsilon_g\equiv\Delta - \Delta_g \equiv E\{Y^{(1)}-Y^{(0)}\}-E\{g(S^{(1)})-g(S^{(0)})\}$ with respect to $g$ under the constraint that $\epsilon_g \ge 0$. Since $g$ is not location identifiable, one may constrain the minimization under both $E\{Y^{(0)}-g(S^{(0)})\}=0$ and $\epsilon_g \ge 0$, which leads to the equivalent minimization problem of
$$
\epsilon^2_g=[E\{Y^{(1)}-g(S^{(1)})\}]^2\  s.t.\  E\{Y^{(0)}-g(S^{(0)})\}=0 \ \text{and}\ \epsilon_g \ge 0.
$$
By Jensen's inequality $\epsilon^2_g=[E\{Y^{(1)}-g(S^{(1)})\}]^2 \leq E\{Y^{(1)}-g(S^{(1)})\}^2$. If we can find a $g$ function such that the loss function $E\{Y^{(1)}-g(S^{(1)})\}^2$ is very small, then the loss $\epsilon^2_g$ will automatically be very small. To this end, as an alternative strategy, we minimize
\begin{align}
L(g)=E\{Y^{(1)}-g(S^{(1)})\}^2\ s.t. \ E\{Y^{(0)}-g(S^{(0)})\}=0. \label{eq1}
\end{align}
Note that we have also dropped the constraint that $\epsilon_g\ge 0$ since it is satisfied automatically with the solution to the optimization problem (\ref{eq1}) under minor conditions that can be empirically checked with observed data, will be shown later. However,  $\gopt(s)$ optimizing (\ref{eq1}) is not uniquely identifiable for $s \in D_0=\Omega_0 \setminus \Omega_1$, where $\Omega_a$ denotes the support of $S^{(a)}$ for $a=0,1$. For identifiability, we let $\gopt(s)=m_0(s)+c$ for $s\in D_0$,
where $m_0(s)=E(Y^{(0)}|S^{(0)}=s)$ and $c$ is an unknown constant to be determined.
Under this constraint, we show in Appendix \ref{app-R1-PTEL} that the following $\gopt$ minimizes (\ref{eq1}):
\begin{equation}
\gopt(s)=
\begin{cases}
\begin{array}{ll}
m_1(s) + \lambda\ r(s),& \ s \in \Omega_1=D_c \cup D_1 \\
m_0(s)+c,&  \ s \in D_0
\end{array}
\end{cases}
\label{gopt}
\end{equation}
where $D_c\equiv \Omega_1\cap\Omega_0$, $D_1=\Omega_1 \setminus \Omega_0$, $m_a(s) =  E(Y^{(a)} \mid S^{(a)}=s),$ $f_a(s)=d F_a(s)/ds$ is the conditional density of $S\supa$ with $F_a(s)=P(S\supa \le s)$, $r(s) = f_0(s)/f_1(s)$ is the density ratio,

\begin{eqnarray*}
\lambda &=&\left\{K_2+K_1 r(s^*) \right\}^{-1} \left[\int_{D_c}\Delta_{01}(s)f_0(s)ds+ K_1 \Delta_{01}(s^*) \right],\\
c&=&\left\{K_2+K_1r(s^*) \right\}^{-1}\left[r(s^*) \int_{D_c}\Delta_{01}(s)f_0(s)ds-K_2\Deltazeroone(s^*)\right]
\end{eqnarray*}
with $\Delta_{01}(s) = m_0(s)-m_1(s)$, $K_1=\int_{D_0}f_0(s)ds$, $K_2=\int_{D_c}  r(s) f_0(s) ds$ and $s^*$ being the intersection point of $D_c$ and $D_0$.
When $\Omega_0\subseteq \Omega_1$, $D_0$ is empty, $K_1=0$, and $\gopt$ is reduced to
\begin{equation}
\gopt(s)=m_1(s) + \lambda\ r(s),  \quad \mbox{where $\lambda = K_2 ^{-1}\int_{D_c}\Delta_{01}(s)f_0(s)ds.$ }
\label{goptsimple}
\end{equation}
\begin{remark}
With the aim of predicting $Y$, a natural choice of $\gopt(s)$ for $s \in D_0$ is $m_0(s)$ as in $D_0$, there are only observations from group 0 with the surrogate marker and thus, $m_0(s)=m(s)=E[Y|S=s]$ for $s \in D_0$ is the best prediction function of $S$ for $Y$. However, an additional constant $c$ is needed to make the function $\gopt(s)$ to satisfy the constraint, and, at the same time, to be continuous at the intersection point $s^*$, where
\begin{eqnarray*}
\gopt(s^*)&=&\frac{r(s^*)}{K_2 +K_1 r(s^*)}\left[\int_{D_c}\Delta_{01}(s)f_0(s)ds  + K_1\Delta_{01}(s^*) \right] + m_1(s^*), \\
\end{eqnarray*}
which is well defined even if $f_1(s^*)=0$.
\end{remark}
\begin{remark}
From the forms of $\lambda$ and $c$, it can be seen that if $m_0(s)=m_1(s)=m(s)$ for $s \in D_c$ (a perfect surrogate), then $\lambda=0$, and $\gopt(s)=m(s)$ for the whole domain. Therefore, $\Delta_{\gopt}=E[\gopt(S^{(1)})-\gopt(S^{(0)})]=E[m(S^{(1)})-m(S^{(0)})]=\int m(s) f_1(s) ds-\int m(s) f_0(s) ds=\int m_1(s) f_1(s) ds-\int m_0(s) f_0(s) ds=\Delta$. That is, $\PTE=1$, which is as would be expected for a perfect surrogate.
\end{remark}

\def\rhat{\widehat{r}}

\subsection{Estimating $\gopt$}\label{gestimation}
We propose to estimate $\gopt$ non-parametrically by first estimating $f_a(s)$, $m_a(s)$ and $\lambda$ as
\begin{eqnarray*}
\fhat_{ a}(s)&=&\ninv_a \sum_{A_i=a} K_{h}(S_{i}-s),\ \mhat_a(s)=\frac{\sum_{A_i=a} K_{h}(S_{i}-s)Y_i  }{\sum_{A_i=a}K_{h}(S_{i}-s) },  \ \Deltahatzeroone(s)=\mhat_0(s)-\mhat_1(s) \\
\lambdahat &=&\left\{\hat{K}_2+\hat{K}_1 \rhat(s^*) \right\}^{-1} \left\{\int_{D_c}\Deltahatzeroone(s)\fhat_0(s)ds+ \hat{K}_1  \Deltahatzeroone(s^*) \right\},\\
\hat{c}&=&\left\{\hat{K}_2+\hat{K}_1\rhat(s^*)\right\}^{-1}\left\{\rhat(s^*) \int_{D_c}\Deltahatzeroone(s)f_0(s)ds-\hat{K}_2\Deltahatzeroone(s^*)\right\},
\end{eqnarray*}
where $\rhat(s) = \fhat_0(s)/\fhat_1(s)$,
$\hat{K}_1=\int_{D_0}\fhat_0(s)ds$, $\hat{K}_2=\int_{D_c}  \rhat(s)\fhat_0(s) ds$, $K_{h}(\cdot)=K(\cdot/h)/h$ is a symmetric kernel function with bandwidth $h = O(n^{-\nu})$, $\nu \in (1/5, 1/2)$.
Based on these quantities, we may construct a plug-in estimate for $\gopt$, denoted by $\ghat$, as follows
 \[ \ghat(s)=\left\{
                \begin{array}{ll}
                  \mhat_1(s) +   \lambdahat  \rhat_0(s),\ s \in D_c \cup D_1 \\
                  \mhat_0(s)+\hat{c},\ s \in D_0.
                \end{array}
              \right. \]
In Appendix \ref{ghat}, we show that $(nh)^{\half}\{\ghat(s) - \gopt(s)\}$ converges in distribution to a normal distribution with mean 0 and variance-covariance $\bSigma^2(s)$.

The resulting PTE for  $\gopt(S)$ can be obtained as $\PTE\subgopt = \Delta\subgopt/\Delta$ and estimated as 
$$\PTEhat\subghat = \Deltahat\subghat/\Deltahat,$$
where $\Deltahat  = \muhat_1 - \muhat_0, \Deltahat_{g} = \muhat_{g,1} - \muhat_{g,0}$, $\muhat_a  = \ninv_a \sum_{i=1}^n I(A_i =a)Y_{i}$, $n_a = \sum_{i=1}^n  I(A_i=a)$, $\muhat_{g,a} =\ninv_a \sum_{i=1}^n  I(A_i=a) g(S_i).$ With respect to PTE, \cite{parast2017evaluating} proposed a class of surrogacy measures based on the PTE to evaluate a surrogate marker, $\PTE_L$, indexed by a reference distribution of the surrogate marker. We show in Appendix \ref{PTE-PTEL} that $\PTE\subgopt$ is approximately equivalent to $\PTE_L$ with a particular reference distribution uniquely defined by $\gopt(\cdot)$ and $\Delta_{\gopt(S)}.$ In addition, this $\PTE\subgopt$ only requires the following conditions (C1) and (C2) to guarantee that $\PTE\subgopt$ is between 0 and 1.
\begin{enumerate}
\item[] (C1)  $\Sbb_1(u)\ge \Sbb_0(u)$ for all $u$,
\item[] (C2) $\Mbb_1(u) \ge \Mbb_0(u)$ for all $u$ in the common support of $\gopt(S\supone)$ and $\gopt(S\supzero)$,
\end{enumerate}
where $\Sbb_a(u)=P\{g(S\supa) > u\mid A=a\}$, $\Mbb_a(u)=E\{Y\supa \mid g(S\supa)=u\}$, for $a=0,1$. These assumptions can be verified based on the observed data and are more likely to hold as $\gopt(S)$ is chosen to be close to $Y$, compared with the four assumptions in \cite{parast2017evaluating}.

\section{Evaluating Surrogacy Using Relative Power \label{methods}}

\subsection{Relative Power Measure \label{resection}}
Our goal is to evaluate the surrogacy of $S$ for the primary outcome $Y$. 
For any $g$ such that $0 \le \Delta_g \le \Delta$, such as $\gopt$ in Section \ref{optimal}, it would be valid to test for the treatment effect $H_0: \Delta = 0$ based on $\Delta_g$.
We propose to quantify the surrogacy of $g(S)$ based on the extent to which the estimated $\Delta_{g}$ can be used to detect the target treatment effect $\Delta$. To this end, consider a pair of regular asymptotically normal estimators $\widehat{\Delta}$ and $\widehat{\Delta}_g$ for $\Delta$ and $\Delta_g$ such that
$$
\nhalf(\Deltahat-\Delta) \to N(0, \sigma^2), \quad \mbox{and}\quad \nhalf(\Deltahat_g - \Delta_g) \to N(0, \sigma_g^2).
$$
Then we may define the effect sizes for $Y$  and $g(S)$ as $\Delta/\sigma$ and $\Delta_g/\sigma_g$, which directly
indicate the potential power of a study in detecting treatment difference $H_0: \Delta=0$ using $Y$ versus using $g(S)$ with a given sample size $\bar{n}$. Thus, we propose to measure the surrogacy of $g(S)$ based on the relative power (RP):
$$
 \RP_g(\bar{n}):= \RP_g(\bar{n},\bar{n}),\ \text{where}\ \RP_g(n_1,n_2)= \frac{\Psc(\Delta_{g}/\sigma_g,n_1)}{\Psc(\Delta/\sigma,n_2)}  ,$$
{where} $\Psc(\Delta_{g}/\sigma_g,n_1)=1-\Phi(1.96-\sqrt{n_1}\ \Delta_g/\sigma_g),\  \Psc(\Delta/\sigma,n_2)=1-\Phi(1.96-\sqrt{n_2}\ \Delta/\sigma),$ the testing powers based on $g(S)$ and $Y$, respectively.
A good surrogate marker will have $\RP_g(\bar{n})$ close to or great than 1 while $\RP_g(\bar{n})$ being much less than 1 would indicate a poor surrogate. Importantly, while $\PTE_g \equiv  \Delta_{g} /\Delta \le 1$ is true with the class of $g$ of interest, it is not necessarily the case that $\RP_g(\bar{n}) \leq 1$. If the variance of $\widehat{\Delta}_{g}$ is sufficiently smaller than that of $\widehat{\Delta},$ $\RP_g(\bar{n})$ may be larger than 1, indicating greater power and efficiency when the effect size is calculated using the surrogate information due to the reduction in variation. Compared with $\PTE_g$, $\RP_g(\bar{n})$ considers variation in estimating $\Delta_{g}$ and provides more direct information on the power of the study if $g(S)$ is used instead of $Y$. We examine both $\RP_g(\bar{n})$ and $\PTE_g$ in our numerical studies.


\def\ninv{n^{-1}}

\subsection{Estimation of RP}

To estimate $\RP_g(\bar{n})$ for a given $g$, we estimate $\Delta$ and $\Delta_g,$ respectively, by
$$\Deltahat  = \muhat_1 - \muhat_0  \quad \mbox{and} \quad \Deltahat_{g} = \muhat_{g,1} - \muhat_{g,0},$$
where.
In Appendix \ref{RP}, we show that  $\sqrt{n}(\Deltahat -{\Delta})$ and $\sqrt{n}(\Deltahat_g -\Delta_g)$ respectively converge in distribution to $N(0, \sigma^2)$ and $N(0, \sigma_g^2),$ where $\sigma^2=E\{\psi^2_{i}\}$ and $\sigma_g^2=E\{\psi^2_{g,i}\}$ can be respectively estimated by $\sigmahat^2 = \ninv \sumin \psihat_{i}^2$ and $\sigmahat_g^2 = \ninv\sumin \psihat_{g,i}^2$.  $\psi_{i}$, $\psi_{g, i}$, $\psihat_{i}$ and $\psihat_{g,i}$ are influence functions and their estimates.  Their rigourous definitions are given in Appendix \ref{RP}. With these estimators, we may construct a plug-in estimator for $\RP_g(\bar{n})$ as
$$
\RPhat_g(\bar{n}):=\RPhat_g(\bar{n},\bar{n})\ \text{where}\ \RPhat_g(n_1,n_2)= \frac{\Psc(\Deltahat_{g}/\sigmahat_g,n_1)} {\Psc(\Deltahat/\sigmahat,n_2)}.
$$

To assess the variability of $\RPhat_g(n_1,n_2)$, one can show that $\sqrt{n}\{\RPhat_g(n_1,n_2)-\RP_g(n_1,n_2)\}$ converges in distribution to a zero-mean normal distribution with variance $\sigma\subRPg^2(n_1,n_2)$ based on the weak convergence of the random vector
$ \sqrt{n}(\Deltahat -{\Delta}, \Deltahat_g -{\Delta}_g , \sigmahat^2-\sigma^2, \sigmahat_{g}^2-\sigma^2_g)\trans.$
In practice, we may approximate $\sigma\subRPg^2(n_1,n_2)$ via the standard perturbation resampling procedures (\cite*[e.g.]{park2003est,cai2005semiparametric}).

With $\gopt(\cdot)$ estimated as $\ghat(\cdot)$, we may estimate $\Delta_{\gopt}$ as $\Deltahat_{\ghat}$ and $\sigma_{\gopt}^2$ as $\sigmahat_{\ghat}^2 = \ninv\sumin \psihat_{\ghat,i}^2$. A plug-in estimate for  $\RP\subgopt(\bar{n})$ can be constructed accordingly, denoted as $\RPhat\subghat(\bar{n})$, whose asymptotic variance can be estimated by perturbation resampling procedures similarly.

\subsection{Using RP to Design a Future Trial \label{futuresection}}
With a surrogate marker identified in an existing trial (Phase 2 trial), it is possible to use our estimate of $\RP$ to inform the design of a new trial (Phase 3 trial) wherein one would use the treatment effect on the surrogate information to predict or test for the treatment effect on the primary outcome. We assume the transportability of $\Delta_{g}/\sigma_g$ between the existing trial and the future trial, which is generally reasonable in the Phase 2 trial and Phase 3 trial setting since these trials usually have the same inclusion-exclusion criteria. 
Under this assumption,  we may consider relative power between a future trial and an existing trial as:
\begin{equation}
{\RP}_g(n^*,\bar{n}) = \frac{\Psc(\Delta_{g}/\sigma_g,n^*)}{\Psc(\Delta/\sigma,\bar{n})},\label{neq}
\end{equation}
where $n^*$ is the sample size in the future trial. ${\RP}_g(n^*,\bar{n})$ can be interpreted approximately as the power ratio of
$$\Delta_{g}/\mbox{se}(\widehat{\Delta}^*_g)\ \text{vs.}\ \Delta/\mbox{se}(\widehat{\Delta}),$$
where $\widehat{\Delta}_{g}^*$ is the estimator of $\Delta_g$ in the future trial with sample size $n^*$ and $\widehat{\Delta}$ is the estimator of $\Delta$ in the existing trial with sample size $\bar{n}$. Of course, (\ref{neq}) can be re-written such that we can use the expression to determine the needed sample size $n^*$ for the future trial given a desired ${\RP}_g(n^*,\bar{n})$ in the future trial.

Alternatively, one could consider selecting $n^*$ such that it is ensured that the lower bound of the one-sided $100(1-\alpha)\%$ confidence interval (CI) for ${\RP}_g(n^*,\bar{n})$ exceeds a desired threshold value $\kappa.$ 



\section{Final Estimation and Inference for $\RP$ with Estimated $\gopt(\cdot)$ \label{fisection}}
Using the same dataset to estimate both $\gopt$ and its corresponding $\RP(\bar{n})=\RP_{\gopt}(\bar{n})$ may lead to  overfitting bias as in standard prediction settings. Therefore, we employ cross-validation (CV) wherein we split the data randomly into two subsets and estimate $\gopt$ with one subset, and estimate $\RP_g(\bar{n})$ given $g$ using a separate subset.

Specifically, denote $\Isc_k$ and $\Isc_{-k} = \{1,...,n\} \setminus \Isc_k,\ k = 1,..., K,$ be a random partition of the index set $\{1,...,n\}$ of equal sizes,  and let $\Dscr\subIsc = \{\bD_i, i \in \Isc\}$.
Let $\ghat\subIsc$ denote $\gopt$ estimated based on $\Dscr\subIsc$. Given $\ghat\subIsck$, $\RP_{\gopt}(\bar{n})$ is estimated using data in $\Dscr\subIscnk$, and denoted by $\RPhat_{\ghat\subIsck}\supnk(\bar{n})$. Then, we define the CV-based estimator of $\RP\subgopt(\bar{n})$ as
$$\RPhat\subcv(\bar{n})= K^{-1}\sum_{k=1}^K \RPhat_{\ghat\subIsck}\supnk(\bar{n}).$$

The consistency of $\ghat\subIsck$ to $\gopt$ and that of $\RPhat_{g}\supnk(\bar{n})$ to $\RP_{g}(\bar{n})$ guarantee the consistency of $\RPhat\subcv(\bar{n})$ to $\RP(\bar{n})$. The asymptotic distribution of $\RPhat\subcv(\bar{n})-\RP(\bar{n})$ can be obtained from the asymptotic expansions of $\ghat\subIsck-\gopt$ and $\RPhat_{g}\supnk(\bar{n})-\RP_{g}(\bar{n})$. Specifically, when $h = O(n^{-\nu})$ with $\nu \in (1/4,1/2)$,
$$
\nhalf\{\RPhat\subcv(\bar{n}) - \RP\subgopt(\bar{n})\} = \nnhalf  \sumin \psi\subRPgopti(\bar{n}) + o_p(1) ,
$$
which converges in distribution to a normal with mean 0 and variance $\tau\subRPgopt^2(\bar{n}) = E\{\psi\subRPgopti ^2(\bar{n})\}$. Similar to $\sigma\subRPg(\bar{n})$, it is difficult to construct explicit estimation of $\tau\subRPgopt^2(\bar{n}) $ and we instead employ resampling methods.
Estimation and inference for $\PTE=\PTE_{\gopt}$, whose estimate we denote as  $\PTEhat\subghat$, can be derived similarly.

\def\subInd{_{\scriptscriptstyle \sf Ind}}

\section{Simulation studies}

\subsection{Simulation Goals}
We conducted simulation studies to: (1) evaluate the finite sample performance of the proposed estimation and inference procedures for $\RP(\bar{n}), \bar{n}=50, 100, 150,$ with respect to bias, accuracy of standard error estimates, and coverage probabilities in a variety of settings, (2) compare estimates of $\RP(\bar{n})$ and PTE, and (3) compare PTE of our proposed optimal transformation with existing PTE methods. Specifically, for comparison of PTEs, we include PTE estimate from the methods of  (i)  {\cite{wang2020model}}, denoted as $\PTE_{W_{2020}}$; (ii) \cite{parast2016robust}, denoted as $\PTE_L$; (iii) \cite{wang2002measure}, denoted as $\PTE_W$; and (iv) \cite{freedman1992statistical}, denoted as $\PTE_F$.

\subsection{Simulation Setup}
We examined five simulation settings; settings were selected in an effort to examine settings with varying surrogate strength (e.g. weak vs. moderate vs. strong surrogate) and settings that violate certain assumptions required by existing comparator methods. Throughout, we let $n=2000$, variances were estimated using perturbation resampling, a normal density was used for the kernel function, and we chose the bandwidth $h=h_{opt}n^{-c_0}$ with $c_0=0.06$ to ensure under-smoothing needed for $\RP(\bar{n})$ estimation, where $h_{opt}$ is obtained using the procedure of \cite{scott1992multivariate}.  For settings $k= 1, 2, 3,$ we generate
\begin{alignat*}{2}
S\supone&\sim Gamma(shape=a^{(1)}_k, scale=b^{(1)}_k), &\quad& S\supzero \sim Gamma(shape=a^{(0)}_k, scale=b^{(0)}_k),\\
Y\supone&=I\{E^{(1)}/(0.2 G_k\supone(S^{(1)}))>t\}, &\quad& Y\supzero=I\{E^{(0)}/(0.2+0.22G_k\supzero(S^{(0)}))>t\},
\end{alignat*}
where $E\supzero$ and $E\supone$ follow the unit exponential distribution, and we let\\
\[
\begin{matrix*}[l]
 a^{(1)}_1 = 2, & b^{(1)}_1 = 2, & a^{(0)}_1 = 9, & b^{(0)}_1 = 0.5, & G_1\supone(s)=s,         &  G_1\supzero(s)&=s;\\
 a^{(1)}_2 = 2, & b^{(1)}_2 = 2, & a^{(0)}_2 = 9, & b^{(0)}_2 = 0.5, & G_2\supone(s)=s-3\log(s),&  G_2\supzero(s)&=3;\\
 a^{(1)}_3 = 5, & b^{(1)}_3 = 1, & a^{(0)}_3 = 9, & b^{(0)}_3 = 0.5, & G_3\supone(s)=s/2,       &  G_3\supzero(s)&=9/11+s.
\end{matrix*}
\]

In setting (4), $S\supone \sim$ Uniform$(1,3)$, $S\supzero \sim$ Uniform$(2,4)$, $G_5\supone(s)=G_5\supzero(s)=s$, and $Y^{(1)}, Y^{(0)}$ are generated as above. In setting (5), we generated $$\left[\begin{matrix} S^{(1)} \\ S^{(0)} \end{matrix}\right] \sim N\left(\left[\begin{matrix} 5 \\ 5 \end{matrix} \right], \left[\begin{matrix} 2 & 1 \\ 1 & 1 \end{matrix} \right] \right),$$
and $Y^{(1)}$ and $Y^{(0)}$ from
$$P(Y^{(1)} \mid S^{(0)}, S^{(1)})=\exp\{-1-0.1{S^{(1)}}^2 \},\ \mbox{and}\ P(Y^{(0)} \mid S^{(0)}, S^{(1)})=\exp\{ - 4 -0.1 {S^{(0)}}^2\}.$$
In setting (1), all assumptions required by \cite{parast2016robust} are satisfied. However, in settings (2), the effect of $S$ on $Y$ is non-monotone and in settings (3) and (4), $S\supzero$ and $S\supone$ have rather different supports; thus, in these settings, the assumptions required by \cite{parast2016robust} do not hold. The working independence assumption of \cite{wang2020model} holds in settings (1)-(4) but not in setting (5).

%
%

\subsection{Simulation Results}
Simulation results are shown in Table \ref{tab-pte} and Table \ref{tab-re}, for PTE and $\RP$, respectively. All results are summarized based on 500 replications for each setting.  Across all settings, the point estimates for our proposed $\RP$ measure using $\gopt$ have negligible bias, estimated standard errors are close to the empirical standard errors, and coverage probabilities of the confidence intervals are close to their nominal level $0.95$. Similar results are observed for the PTE estimate using our proposed $\gopt$. With respect to comparing $\RP$ and PTE, in setting (1) and (4), where the estimates of $\PTE\subgopt$ are relatively higher (above 0.5) than other settings, the estimates of $\RP\subgopt(\bar{n})$ are above 1, so PTE and $\RP$ are consistent in indicating the surrogacy of a surrogate marker. This also suggests that although the estimated $\Delta\subgopt$ is slightly smaller compared to $\Delta$, the variation of $\Deltahat\subgopt$ is substantially smaller than the corresponding variation of $\Deltahat$, leading to higher power if the study were to be based on $\gopt(S)$ rather than the outcome $Y$ itself. This illustrates the advantage of using $\RP$ for quantifying surrogacy since it is more closely tied to study power and effect size compared to PTE alone.

Table \ref{tab-pte}  also summarizes the results of other PTE estimators. Across all settings, the methods of \cite{wang2002measure} and \cite{freedman1992statistical} misspecify the underlying model and as a result, $\PTE_W$ and $\PTE_F$ estimates differ substantially from the nonparametric estimates of $\PTE$ (using $\gopt$), $\PTE_{W_{2020}}$ and $\PTE_L$. For setting (2), where we have introduced a deviation from the monotone increasing assumption for $E(Y \mid S=s)$, we observe that except for our proposed PTE and $\PTE_{W_{2020}}$ estimates, the other methods yield PTE estimates negative or close to zero. This is due to the fact that the monotone assumption fails in this case and our proposed PTE and $\PTE_{W_{2020}}$ evaluate the PTE for $\gopt(S)$ rather than $S$, thus demonstrating the robustness of the proposed PTE and $\PTE_{W_{2020}}$. In setting (3), $\PTE_L$, $\PTE_W$ and $\PTE_F$ all fail with their estimates being negative. This may be due to the supports of the treatment and control groups being quite different. In contrast, the proposed PTE and $\PTE_{W_{2020}}$ perform well here. In setting (4), both our proposed PTE and $\PTE_{W_{2020}}$ estimates identify this setting as one with strong surrogacy while the comparison methods fail to do so. Across all settings, the proposed PTE estimates are comparable or a little bit larger than corresponding estimates of $\PTE_{W_{2020}}$, so both estimators are relatively robust and comparable.


\section{Application to the Diabetes Prevention Program Study}

To illustrate our proposed $\RP$ measure, we apply our procedures to the Diabetes Prevention Program (DPP) study which was a randomized trial investigating the effect of several prevention strategies for reducing the risk of type 2 diabetes (T2D) among high risk individuals with pre-diabetes (\cite*{diabetes1999diabetes}, \citeyear{DPPOS_NEJM}). DPP data are publicly available through the the
National Institute of Diabetes and Digestive and Kidney Diseases Central Repository. The participants were randomized to one of four treatment groups: placebo, lifestyle intervention, metformin and troglitazone. The primary endpoint of the trial was time to T2D onset and the participants were followed up to 5 years with a mean follow up of 2.8 years. Both lifestyle intervention and metformin were shown to significantly reduce T2D risk among participants.

For illustration, we focused on the comparison of the lifestyle intervention group ($n_1=1007$) versus the placebo group ($n_0 = 1010$) with respect to diabetes risk at $t =1, 2, 3, 4$ years. Our goal is to investigate to what extent surrogate information on hemoglobin A1C (HbA1c) or fasting glucose at $t_0 =0.5$ years (i.e., 6 months), can be used to predict treatment effect on diabetes risk at $t=1, 2, 3$ or $4$ years. Only 10 patients developed T2D before $t_0$ and were excluded from this analyses. We evaluate the surrogacy potential of these markers based on the proposed $\RP$ measure primary, and also calculate PTE for comparison.

Results are shown in Table \ref{examplepte} and Table \ref{examplere}. Examining $\RP$ first, for both HbA1c and glucose, $\RP$ is highest when $t=1$ and generally decreases as $t$ gets further from $t_0=0.5$. For comparison, we provide the other PTE estimates as well; the PTE estimate with the proposed $\gopt$ is generally larger than or comparable to corresponding estimates of $\PTE_{W_{2020}}$, $\PTE_L$, $\PTE_W$ and $\PTE_F$, which is similar to what was observed in simulations meaning that the proposed transformed surrogate explains a larger proportion of the treatment effect on the outcome than the untransformed surrogate. In addition, using either PTE or $\RP$, fasting glucose appears to be a stronger surrogate compared to HbA1C.

To illustrate how these estimates can be used to design a future trail, consider the estimated $\RP(n^*, 50)$ in (\ref{neq}) for the primary outcome at $t = 2$ for fasting glucose. To ensures a $95\%$ lower bound of $\RPhat(n^*, 50)$ above 1, we obtain a needed sample size $n^*\geq 50$. This suggests that with a future sample size $n^*\geq 50$, the power of a future 0.5-year trial based on $\gopt(S)$ could be at least as high as the power of the DPP study with sample size 50 based on the diabetes onset information collected up to 2 years.

\section{Discussion}

We derive an optimal transformation of the surrogate marker such that we avoid the requirement of a working independence assumption in \cite{wang2020model}. Our methods has the advantage of both being model-free and requiring flexible assumptions about the surrogate marker distribution and its relationship with the outcome. Numerical studies show good performance of this optimal transformation. Both the proposed optimal transformation and the optimal transformation of \cite{wang2020model} are robust to various scenarios and have comparable performances. So the proposed one is a good alternative to the one in \cite{wang2020model} if one is not confident in the validity of the working independence assumption for a specific dataset under analysis. 

We propose a relative power measure to quantify the utility of a potential surrogate marker which is measured either earlier than the primary outcome or with less burden/cost compared to the primary outcome. Unlike the commonly used proportion of treatment effect explained measure, the $\RP$ measure provides a direct link to the expected power of subsequent phase trials and can be used to inform their design. Specifically, it directly reflects the expected gain or loss in power when considering the use of a surrogate marker in a future trial relative relying on the primary outcome.  Through the calculation of a sample size, and if desired, a confidence interval lower bound, actionable information to determine needed study size and duration can be obtained. We have provided a nonparametric inference approach for the optimal transformation and the corresponding $\RP$, which demonstrated good finite sample performance. 

Importantly, the ability to calculate a sample size to inform the design of a future trial relies on the assumption of transportability of the quantity $\Delta_{g}/\sigma_g$ from the existing trial to a future trial. This is reasonable for different phases of trials as these trials often use parallel inclusion criteria of participants. But using surrogate to inform different future studies needs caution. According to our knowledge, the transportability is unavoidable in studying surrogate markers. We choose to assume the transportability of $\Delta_{g}/\sigma_g$ instead of, for example, the complete joint distribution of outcome and surrogate marker. Transportability of information learned about a surrogate marker in a previous study is a complex and interesting issue and is an active area of research (\cite*{wang2020model,price2018estimation,athey2016estimating}). Of course, the ultimate goal underlying surrogate marker research is that if they can be identified, they can be used in  future trials, and reduce follow-up time and costs, but successfully achieving this goal strongly relies on the assumption of transportability. Violations of this transportability assumption could have important consequences and future work in this area is warranted.

Our work has some limitations. Given our nonparametric estimation approach, we require a relatively large sample size such that the kernel smoothing will behave properly. Our methods would likely not be a reasonable option for studies with a very small sample size and in those cases, a parametric approach may need to be considered. In addition, we can address the issue of drop-out, censoring, or staggered entry into the study similarly to \cite{wang2021quantifying}. Each of these issues would introduce additional complexities that warrant future work. Lastly, we focus on evaluating and using a single surrogate marker. Often, studies have multiple potential surrogate markers and/or a surrogate marker measured repeatedly over time i.e., a longitudinal marker (\cite*{wang2022robust}). While methods have been developed to evaluate surrogate in these settings, this area of research would benefit from further development of methods that address the issue of how to design future clinical trial studies that would use such markers to replace the primary outcome (\cite*{parast2020evaluating,athey2019surrogate,agniel2020evaluation}).

The DPP data used in this paper is not available to the public. The authors can provide information on application for the dataset.

\clearpage

\newpage
\begin{table}[ht]
\begin{center}
\end{center}
\centering
\begin{tabular}{r rrrr| rr| rr| rr| rr }  \hline
  \multicolumn{5}{c|}{Proposed} & \multicolumn{2}{c|}{$\PTE_{W2020}$}& \multicolumn{2}{c|}{$\PTE_L$}& \multicolumn{2}{c|}{$\PTE_W$} & \multicolumn{2}{c}{$\PTE_F$} \\
  & True & Est & ESE$_{\mbox{\tiny ASE}}$ & CP & Est & ESE & Est & ESE & Est & ESE& Est & ESE \\  \hline
\multirow{1}{*}{(1)} &
  .657 &.666 &.073$_{.074}$& .956 & \multirow{ 1}{*}{.616}  & \multirow{1}{*}{.060} &\multirow{1}{*}{.374} & \multirow{1}{*}{.078} & \multirow{1}{*}{.195} & \multirow{1}{*}{ .043}& \multirow{1}{*}{ .189}& \multirow{1}{*}{.041} \\
 \hline
\multirow{1}{*}{(2)} &
  .188 &.207 &.057$_{.068}$ &.968 & \multirow{1}{*}{.140}  & \multirow{1}{*}{.049} &\multirow{1}{*}{-.226} & \multirow{1}{*}{.060} & \multirow{1}{*}{.075} & \multirow{1}{*}{.020}& \multirow{1}{*}{.059} & \multirow{1}{*}{.016}  \\
\hline
\multirow{1}{*}{(3)} &
  .095 &.092 &.023$_{.027}$ &.972 & \multirow{1}{*}{.077}  & \multirow{1}{*}{.015} &\multirow{1}{*}{-.042} & \multirow{1}{*}{.013} & \multirow{1}{*}{-.049} & \multirow{1}{*}{.011}& \multirow{1}{*}{-.037} & \multirow{1}{*}{.008} \\
\hline
\multirow{1}{*}{(4)} &
   .772 &.794& .056$_{.062}$ &.942 & \multirow{1}{*}{.806}  & \multirow{1}{*}{.033} &\multirow{1}{*}{.382} & \multirow{1}{*}{.280} & \multirow{1}{*}{.546} & \multirow{1}{*}{.118}& \multirow{1}{*}{.463} & \multirow{1}{*}{.086}  \\
 \hline
\multirow{1}{*}{(5)} &
  .301 &.308 &.080$_{.101}$ &.986 & \multirow{1}{*}{.244}  & \multirow{1}{*}{.057} &\multirow{1}{*}{.177} & \multirow{1}{*}{.065} & \multirow{1}{*}{.001} & \multirow{1}{*}{.039}& \multirow{1}{*}{.001} & \multirow{1}{*}{.027}  \\
 \hline
\end{tabular}
\caption{Estimates (Est) of $\PTE$ (using our proposed $\gopt$), $\PTE_{W2020}$, $\PTE_L$, $\PTE_W$, and $\PTE_F$ along with their empirical standard errors (ESE) under settings (1)-(5); for PTE estimates using our proposed $\gopt$, we also present the averages of the estimate standard errors (ASE, shown in subscript) along with the empirical coverage probabilities (CP) of the 95\% confidence intervals. } \label{tab-pte}
\end{table} 

\begin{table}[ht]
\begin{center}
\end{center}
\centering
\begin{tabular}{rr rrrr  }  \hline
& & True & Est & ESE$_{\mbox{\tiny ASE}}$ & CP   \\  \hline
\multirow{2}{*}{(1)} &
 $\RP(50)$ & 2.005 &2.117 &.415$_{.409}$ &.934     \\
& $\RP(100)$ & 1.710 &1.792 &.360$_{.372}$ &.952     \\
& $\RP(150)$ & 1.432 &1.501 &.277$_{.304}$ &.964   \\
 \hline
\multirow{2}{*}{(2)} &
 $\RP(50)$ & .631& .707& .253$_{.320}$ &.974    \\
& $\RP(100)$ & .615 &.679 &.251$_{.324}$ &.980     \\
& $\RP(150)$ & .631& .673 &.231$_{.297}$ &.976     \\
\hline
\multirow{2}{*}{(3)} &
 $\RP(50)$ & .216& .200 &.054$_{.059}$ &.919     \\
& $\RP(100)$ & .378 &.344 &.099$_{.108}$ &.913     \\
& $\RP(150)$ & .521& .469& .127$_{.142}$ &.921     \\
\hline
\multirow{2}{*}{(4)} &
 $\RP(50)$ & 2.231& 2.337& .374$_{.400}$ &.962     \\
& $\RP(100)$ & 1.361 &1.421 &.174$_{.199}$ &.972    \\
& $\RP(150)$ & 1.128 &1.167 &.095$_{.115}$ &.984     \\
 \hline
\multirow{2}{*}{(5)} &
  $\RP(50)$ & .783& .737 &.184$_{.203}$ &.938     \\
& $\RP(100)$ & .759 &.705& .210$_{.235}$ &.938     \\
& $\RP(150)$ & .761 &.700& .215$_{.241}$ &.932    \\
 \hline
\end{tabular}
\caption{Estimates (Est) of $\RP(\bar{n})$ (using our proposed $\gopt$), with their empirical standard errors (ESE), the averages of the estimate standard errors (ASE, shown in subscript) and the empirical coverage probabilities (CP) of the 95\% confidence intervals under settings (1)-(5). } \label{tab-re}
\end{table} 

\newpage
\begin{table}[ht]
\centering
HBA1c \\
\begin{tabular}{r|c|c|c|c|c }  \hline
 &  $\PTE $   & $\PTE_{W2020}$  & $\PTE_{L}$  &   $\PTE_{W}$&  $\PTE_{F}$ \\ \hline
 $t=1$  &   .264$_{.066}$  &.240$_{.034}$  &.241$_{.011}$  & .163$_{.007}$      & .195$_{.004}$   \\
 $t=2$  &   .331$_{.076}$ & .250$_{.025}$  & .179$_{.005}$  & .155$_{.003}$  &  .208$_{.003}$    \\
 $t=3$  &  .295$_{.055}$ & .248$_{.020}$ &.186$_{.004}$  &.137$_{.002}$  &  .176$_{.002}$    \\
 $t=4$  &  .286$_{.060}$  &.240$_{.021}$ & .169$_{.004}$  &  .139$_{.003}$    &  .175$_{.002}$     \\
  \hline
\end{tabular}
\\
\bigskip
\centering
Fasting glucose \\
\begin{tabular}{r|c|c|c|c|c }  \hline
 &  $\PTE $   & $\PTE_{W2020}$  & $\PTE_{L}$  &   $\PTE_{W}$&  $\PTE_{F}$ \\ \hline
 $t=1$  &   .558$_{.067}$ & .475$_{.044}$ & .337$_{.018}$ &  .267$_{.015}$&  .478$_{.014}$   \\
 $t=2$  &   .570$_{.067}$ & .535$_{.035}$ & .603$_{.021}$  &  .449$_{.014}$   & .536$_{.013}$     \\
 $t=3$  & .564$_{.061}$ & .515$_{.029}$ & .495$_{.011}$ &  .382$_{.008 }$  &  .478$_{.008}$    \\
 $t=4$  & .576$_{.068 }$& .521$_{.030}$ & .479$_{.012}$   &  .377$_{.010}$    &  .481$_{.010}$   \\
  \hline
\end{tabular}
\caption{Estimates of $\PTE$ using the proposed $\gopt$, and $\PTE_{W2020}$, $\PTE_{L}$,  $\PTE_{W}$, $\PTE_{F}$, along with the estimated standard errors (shown in subscript). } \label{examplepte}
\end{table}

\begin{table}[ht]
\centering
HBA1c \\
\begin{tabular}{r|c|c|c }  \hline
 &$\RP(50)$&  $\RP(100)$   & $\RP(150)$    \\ \hline
 $t=1$  & 1.054$_{.305}$ & 1.065$_{.369}$  &1.067$_{.379}$      \\
 $t=2$  & .921$_{.268}$ & .924$_{.309}$ & .930$_{.306}$     \\
 $t=3$  &.702$_{.185}$ & .694$_{.196}$ & .717$_{.188}$      \\
 $t=4$  &.770$_{.227}$ & .758$_{.246}$  &.772$_{.239}$      \\
  \hline
\end{tabular}
\\
\bigskip
\centering
Fasting glucose \\
\begin{tabular}{r|c|c|c }  \hline
 &$\RP(50)$&  $\RP(100)$   & $\RP(150)$    \\ \hline
 $t=1$  & 1.756$_{.375}$ & 1.808$_{.418}$ & 1.728$_{.397}$     \\
 $t=2$  & 1.486$_{.294}$ & 1.488$_{.304}$ & 1.417$_{.272}$       \\
 $t=3$  &1.296$_{.248}$  &1.267$_{.240}$ & 1.208$_{.204}$     \\
 $t=4$  &1.466$_{.297}$ &1.431$_{.295 }$& 1.341$_{.254}$   \\
  \hline
\end{tabular}
\caption{Estimates of $\RP$ using the proposed $\gopt$ along with the estimated standard errors (shown in subscript). } \label{examplere}
\end{table}

\clearpage
\appendix
\begin{center}
{\large \bf Supplementary Materials} \vspace{.2in}

%
%
%

\end{center}

\def\gtildeopt{\widetilde{g}_{\mbox{\tiny opt}}}

\section{Influence functions of $\Deltahat, \Deltahat_g, \sigmahat $ and $\sigmahat_g ^2$}\label{RP}

We derive asymptotic distributions for $\Deltahat $ and $\Deltahat_g $. To this end, we first write
\begin{align*}
& \muhat_a -\mu_a   =\frac{\sum_{i=1}^n I(A_i=a)   Y_i }{\sum_{i=1}^n I(A_i=a) }- \mu_a=\frac{\sum_{i=1}^n I(A_i=a) (  {\displaystyle{ Y_i}}{}-  \mu_a) }{\sum_{i=1}^n I(A_i=a) }:= n^{-1}\sum_{i=1}^n \psi_{a,i}  +o_p(n^{-1/2}),
\end{align*}
where $\psi_{a,i}=2 I(A_i=a) ( {\displaystyle{ Y_i}}{}-  \mu_a)$.
It follows that
\begin{align*}
\sqrt{n}(\Deltahat -\Delta)=n^{-1/2}\sum_{i=1}^n (\psi_{1,i} -\psi_{0,i} ) +o_p(1):=n^{-1/2}\sum_{i=1}^n \psi_{i}  +o_p(1).
\end{align*}
By the central limit theorem we have that $\sqrt{n}(\Deltahat -\Delta )$ converges in distribution to a normal distribution $N(0, \sigma^2 )$ with $\sigma^2=E[ \psi_{i}^2 ]. $

Similarly, we have
\begin{align*}
& \muhat_{g,a} -\mu_{g,a}   =\frac{\sum_{A_i=a}  g(S_i) }{\sum_{A_i=a} 1 }- \mu_{g,a} =\frac{\sum_{A_i=a}  \{g(S_i)- \mu_{g,a}\} }{\sum_{A_i=a} 1 }  : =  \ninv \sum_{i=1}^n \psi_{g,a,i}  + o_p(n^{-1/2}).
\end{align*}
It follows that
$$\sqrt{n}(\Deltahat_g -\Delta_g )=n^{-1/2}\sum_{i=1}^n (\psi_{g,1,i} -\psi_{g,0,i} ) +o_p(1):=n^{-1/2}\sum_{i=1}^n \psi_{g,i}  +o_p(1).$$
By the central limit theorem we have that $\sqrt{n}(\Deltahat_g -\Delta_g )$ converges in distribution to a normal distribution $N(0, \sigma_g^2 )$ with $\sigma_g^2 =E[ \psi^2_{g, i} ].$

We next derive estimators for the asymptotic variances $\sigma^2 $ and $\sigma_g^2 $. To this end, we first note that the variance of $\psi_{a,i} $ is
\begin{eqnarray*}
 E \psi^2_{a,i} &=&E [ 4 I(A_i=a) ( Y_i -\mu_a)^2],
\end{eqnarray*}
which can be estimated by
\begin{eqnarray*}
\hat{\Sigma}_{a} &=&n^{-1}\sum_{i=1}^n 4 I(A_i=a) (Y_i -\hat{\mu}_a)^2.
\end{eqnarray*}
Therefore, the asymptotic variance of $\sqrt{n}(\Deltahat -\Delta )$, $\sigma^2 $, can be estimated by $\sigmahat^2 := \hat{\Sigma} :=\hat{\Sigma}_1 +\hat{\Sigma}_0 .$

It follows from the above formulas that $n^{1/2}(\sigmahat^2 -{\sigma}^2 )$ can be written as the form $$n^{1/2}(\sigmahat^2 -{\sigma}^2 )=n^{-1/2} \sum_{i=1}^n \psi_{\sigma^2, i} +o_p(1).$$

Similarly, we can get the estimate for the asymptotic variance of $\sqrt{n}(\Deltahat_g -\Delta_g )$, $\sigmahat^2_g $, with a given $g$, and
$$n^{1/2}(\sigmahat^2_g -{\sigma}_g^2 )=n^{-1} \sum_{i=1}^n \psi_{\sigma_g^2, i} +o_p(1).$$

With the above influence functions for $\sigmahat^2 $ and $\sigmahat^2_g $, the variance estimates for $\sigmahat^2 $ and $\sigmahat^2_g $ can be obtained by perturbation resampling method.

\section{Derivations of the optimal $g$} \label{app-R1-PTEL}


In this section, we derive the specific form for the optimal transformation function of the surrogate information, $\gopt(\cdot)$.
We aim to solve the following problem for $g$:
\begin{align*}
\min_g L(g)=E\{Y^{(1)}-g(S^{(1)})\}^2, \quad \mbox{given} \quad  E\{Y^{(0)}-g(S^{(0)})\}=0
\end{align*}
with $g(s)=m_0(s)+c, s \in D_0$ and $g(s)$ is continuous. 

Without loss of generality, we assume that $S$ is continuous with conditional densities given $A = a$, $\dot{F}_a(s):=f_a(s)$, with respect to
the Lebesgue measure. Similar arguments as given below can be used to derive $\gopt$ when $S$ is discrete.
It can be shown that
$$ E\{Y^{(1)}-g(S^{(1)})\}^2  \propto E[g^2(S^{(1)})]- 2E[Y^{(1)}g(S^{(1)})] =E[g^2(S^{(1)})]- 2E[m_1(S^{(1)})g(S^{(1)})]. 
$$
And thus the problem is equivalent to finding a function $\gopt(\cdot)$ such that
$$\min_{ g} \frac{1}{2} E[g^2(S^{(1)})]- E[m_1(S^{(1)})g(S^{(1)})] \quad \mbox{given} \quad E [g(S) | A = 0] = \mu_0.$$ Our optimization problem is thus,
$$\min_{g} \int \half g^2(s) f_1(s) ds-  \int m_1(s)g(s) f_1(s) ds \quad \mbox{given that} \int g(s) f_0(s) ds  = \mu_0,$$
which is equivalent to
$$\min_{g} \Lsc( g), \quad \mbox{given that} \quad \Gbb( g) = \mu_0,$$
where we used the functional notation
$$\Lsc( g) = \int \half g^2(s) f_1(s) ds-  \int m_1(s)g(s) f_1(s) ds, \quad \mbox{and} \quad \Gbb( g) =  \int g(s) f_0(s) ds.$$
Taking the Frechet derivatives of the functionals, we have that for all measurable $h$ such that $ \int h^2(s) f_1(s) ds < \infty$,
$$\frac{d}{d g} \bigg [\Lsc(g) - \lambda \Gbb( g)\bigg] (h) = \int \gopt(s) h(s) f_1(s)ds-\int \gopt(s) h(s) m_1(s) f_1(s)ds -  \lambda \int h(s) f_0(s)ds = 0.$$
Setting $h = \delta(s)$, this implies that
$$\gopt(s) = m_1(s)+\lambda f_0(s)/f_1(s)=m_1(s)+\lambda\  r(s), s \in D_c \cup D_1.$$
By the constraint $\int_{D_c} \{m_1(s)+\lambda\ r(s)\}f_0(s) ds+\int_{D_0} \{m_0(s)+c\} f_0(s)ds=\mu_0=\int m_0(s) ds$ and $\gopt(s)$ is continuous at $s^*$, or $m_1(s^*)+\lambda\ r(s^*)=m_0(s^*)+c$, we have
\begin{eqnarray*}
\lambda &=&\left\{K_2+K_1 r(s^*) \right\}^{-1} \int_{D_c} \Delta_{01}(s) f_0(s)ds+ K_1 \left\{K_2+K_1 r(s^*)\right\}^{-1} \! \Delta_{01}(s^*),\\
c&=&\left\{K_2+r(s^*)K_1\right\}^{-1}\left[-K_2 \Delta_{01}(s^*)+r(s^*) \int_{D_c} \Delta_{01}(s) f_0(s)ds\right]
\end{eqnarray*}
with $\Delta_{01}(s) = m_0(s)-m_1(s)$, $K_1=\int_{D_0}f_0(s)ds$, $K_2=\int_{D_c}  f_0^2(s)/f_1(s) ds=\int_{D_c} r(s) f_0(s) ds$.

Finally, the optimal function $\gopt(\cdot)$ can be expressed as
\begin{equation*}
\gopt(s)=
\begin{cases}
&m_1(s) + \lambda\ r(s),\ s \in D_c \cup D_1 \\
&m_0(s)+c,\ s \in D_0.
\end{cases}
\label{gopt}
\end{equation*}

\def\subrte{_{\mbox{\tiny rte}}}
\section{Relationship between $\PTE$ and $\PTE_L$ } \label{PTE-PTEL}

In this section, we show the relationship between our proposed $\PTE$ and the PTE of \cite{parast2016robust}. 
To this end, let $\Delta\subrte$ denote the ``residual treatment effect"  defined in  \cite{parast2016robust} as:
\begin{eqnarray*}
\Delta\subrte&=&\int E(Y\supone-Y\supzero|S\supone=S\supzero=s)d \Fscr(s)
 = \int \{m_1(s) - m_0(s)\} d \Fscr(s),
\end{eqnarray*}
where $\Fscr(\cdot)$ is a reference distribution function. It follows that
\begin{equation}
\Delta_L = \Delta-\Delta\subrte=\int m_1(s)\{d F_1(s)-d \Fscr(s)\}-\int m_0(s)\{d F_0(s)-d \Fscr(s)\}.    \label{def-DL}
\end{equation}
and $\PTE_L=\Delta_{L}/\Delta.$ 

To relate $\Delta_L$ to $\Delta_{\gopt(S)}$, recall that 
\begin{eqnarray*}
\gopt(s)&=& m_1(s) + \lambda f_0(s)/f_1(s), s \in D_c\cup D_1,\\
\gopt(s)&=& m_0(s)+c, s \in D_0,\\
\lambda &=&\left\{K_2+K_1 \frac{f_0(s^*)}{f_1(s^*)} \right\}^{-1}\!\!\!\!\int_{D_c}\{m_0(s)\!-\!m_1(s)\}f_0(s)ds+ K_1 \left\{K_2+K_1 \frac{f_0(s^*)}{f_1(s^*)} \right\}^{-1}\!\!\!\! \{m_0(s^*)\!-\!m_1(s^*)\},\\
c&=&\left\{1+\frac{f_0(s^*)}{f_1(s^*)}\frac{K_1}{K_2} \right\}^{-1}\left[m_1(s^*)-m_0(s^*)+\frac{f_0(s^*)}{f_1(s^*)}\frac{ 1}{K_2} \int_{D_c}\{m_0(s)-m_1(s)\}f_0(s)ds\right]  \text{and}\ \\
\Delta_{\gopt(S)}&=&E\{\gopt(S\supone)-\gopt(S\supzero)\}.
\end{eqnarray*}
Therefore,
\begin{align}
\Delta_{\gopt(S)}&=\int_{D_1} m_1(s)f_1(s)ds+\int_{D_c}\{m_1(s)+\lambda f_0(s)/f_1(s)\} f_1(s)ds -\mu_0 \nonumber \\
&=\int_{D_1 \cup D_c} m_1(s)f_1(s)ds-\int_{D_0 \cup D_c} m_0(s)f_0(s)ds +\int_{D_c}f_0(s)ds \nonumber \\
&\times \left[\left\{K_2+K_1 \frac{f_0(s^*)}{f_1(s^*)} \right\}^{-1}\!\!\!\!\int_{D_c}\{m_0(s)\!-\!m_1(s)\}f_0(s)ds+ K_1 \left\{K_2+K_1 \frac{f_0(s^*)}{f_1(s^*)} \right\}^{-1}\!\!\!\! \{m_0(s^*)\!-\!m_1(s^*)\}\right]\nonumber \\
&= \int_{D_1} m_1(s)f_1(s)ds+\int_{D_c}m_1(s)\left[f_1(s)-f_0(s)\frac{ \int_{D_c}f_0(s)d(s) }{ K_2+K_1 f_0(s^*)/f_1(s^*)  } \right]ds \nonumber \\
&-\int_{D_0 } m_0(s)f_0(s)ds -\int_{D_c}m_0(s) \left[f_0(s)-f_0(s)\frac{ \int_{D_c}f_0(s)d(s) }{K_2+K_1 f_0(s^*)/f_1(s^*) } \right]ds \nonumber \\
&+\int_{D_c}f_0(s)ds\ K_1 \left\{K_2+K_1 \frac{f_0(s^*)}{f_1(s^*)} \right\}^{-1}\!\!\!\! \{m_0(s^*)\!-\!m_1(s^*)\} \nonumber \\
&:=\int m_1(s)\{d F_1(s)-d \Fscr\subnew(s)\}-\int m_0(s)\{d F_0(s)-d \Fscr\subnew(s)\}\nonumber\\
&+\int_{D_c}f_0(s)ds\ K_1 \left\{K_2+K_1 \frac{f_0(s^*)}{f_1(s^*)} \right\}^{-1}\!\!\!\! \{m_0(s^*)\!-\!m_1(s^*)\}\nonumber
\end{align}
where 
\begin{equation}
\Fscr\subnew(s)= \int_{D_c} I(v\leq s) f_0(v) dv  \frac{ \int_{D_c}f_0(s)ds }{K_2+K_1 f_0(s^*)/f_1(s^*) }. \label{fnew}
\end{equation}
If $\Fscr(s)$ in (\ref{def-DL}) is replaced by $\Fscr\subnew(s)$, then $\Delta_{\gopt(S)} = \Delta_L+\int_{D_c}f_0(s)ds\ K_1 \left\{K_2+K_1 \frac{f_0(s^*)}{f_1(s^*)} \right\}^{-1}\!\!\!\! \{m_0(s^*)\!-\!m_1(s^*)\}$; and thus, when $D_0$ is empty ($K_1$=0), or $m_0(s^*)=m_1(s^*)$ we have $\PTE \equiv \PTE_L$ .



\ \

We next show that only assumptions (C1) and (C2) are required to ensure that the proposed $\PTE$ is between 0 and 1.
\begin{eqnarray*}
(\mbox{C1})&& \Sbb_1(u)\ge \Sbb_0(u)\ \text{for all $u$},\\
(\mbox{C2})&& \Mbb_1(u) \ge \Mbb_0(u) \ \text{for all $u$ in the common support of $\gopt(S\supone)$ and $\gopt(S\supzero)$},
\end{eqnarray*}
where $\Sbb_a(u)=P\{\gopt(S\supa) > u\mid A=a\}$, $\Mbb_a(u)=E\{Y\supa \mid \gopt(S\supa)=u\}$, $a=0,1$, which are assumed to be continuous functions. Following arguments given in Appendices \ref{RP} and \ref{app-R1-PTEL},  we have
\begin{align}
\Delta&=E\{Y\supone\}-E\{Y\supzero\}=\int \Mbb_1(u) d \Fbb_1(u)-\int \Mbb_0(u) d \Fbb_0(u) , \nonumber \\
\Delta_{\gopt}&=\int \Mbb_1(u)\{d \Fbb_1(u)\!-\!d \Fbb\subnew(u)\}\!-\!\int \Mbb_0(u)\{d \Fbb_0(u)\!-\!d \Fbb\subnew(u)\} \!+\!H(u^*)\{\Mbb_0(u^*)\!- \!\Mbb_1(u^*)\} \nonumber, \\
\Delta-\Delta_{\gopt}&=\int_{\Dbb_c} \{\Mbb_1(u)- \Mbb_0(u)\}\dot{\Fbb}\subnew(u) du+H(u^*)\{\Mbb_1(u^*)- \Mbb_0(u^*)\} , \label{eq1}
\end{align}
where $H(u^*)$ is a non-negative function of $u^*$, $\Fbb_a(u) = 1-\Sbb_a(u)$, $\dot{\Fbb}_a(u) = d \Fbb_a(u)/du$,
and $\dot{\Fbb}\subnew(u)$ is similarly defined as (\ref{fnew}) but for $\gopt(S)$ instead of $S$ and $\Dbb_c$ is the common support of $\gopt(S\supone)$ and $\gopt(S\supzero)$. It is also straightforward to see from an integration by parts that
\begin{align*}
\Delta_{\gopt(S)}=\int u\ d \Fbb_1(u)-\int u\ d \Fbb_0(u)
=\int \{\Sbb_1(u)-\Sbb_0(u)\}du .
\end{align*}
Thus, from condition (C1), we have $\Delta_{\gopt(S)} \ge 0$. On the other hand, since $\dot{\Fbb}\subnew(u) \ge 0$, we see from (\ref{eq1}) that $\Delta - \Delta_{\gopt(S)} \ge 0$ under condition (C2). It follows that $\PTE \in [0, 1]$ under conditions (C1) and (C2).
Furthermore, $\Delta_{\gopt(S)} = 0$ when $\Delta = 0$.

\section{Asymptotic properties for $\ghat(\cdot)$ } \label{ghat}
Throughout, we assume that $m_a(s), a=0,1$ is continuously differentiable. In addition, we assume that $f_a(s), a=0,1$ is continuously differentiable with finite support. For inference, we require under-smoothing with $h = o_p(n^{-1/5})$ for interval estimation of $\gopt$ and $h=o_p(n^{-1/4})$ for the interval estimation of $\RE$ and $\PTE$. Since $\mhat_a(s)$ and $\fhat_a(s), a=0,1$ are standard kernel estimators, we have that they are consistent w.r.t their true values with rate $(\log n)^{\half}(nh)^{-\half}+ h^2$. It follows immediately that $|\ghat(s)-\gopt(s) | = O_p\{(\log n)^{\half}(nh)^{-\half}+ h^2\}$.

We firstly derive the influence functions for each estimator in Section \ref{gestimation}. The influence functions can be derived following exactly the derivations of $\muhat_{a} -\mu_{a} $ and $\muhat_{g,a} -\muhat_{g,a} $. Direct calculations show that \begin{eqnarray*}
 \fhat_{ a}(s)-f_{a}(s)  &=&\frac{n^{-1 } {\sum_{A_i=a} \{K_{h}(S_{i}-s)-f_a(s)\} }}{n^{-1}\sum_{A_i=a}1  }  \\
&:=&(nh)^{-1}\sum_{i=1}^{n}\phi_{a,i} (s) +o_p\{(nh)^{-1/2}\},\\
\mhat_a(s)-m_a(s) &=& \frac{\sum_{i=1}^{n}I(A_i=a)K_{h}(S_{i}-s)\{Y_i-m_a(s) \}  }{\sum_{i=1}^{n}I(A_i=a) K_{h}(S_{i}-s)  }\\
& := & (nh)^{-1}\sum^{n}_{i=1}\phi_{m,i}(s)+o_p\{(nh)^{-1/2}\},\\
\hat{K}_1-K_1&=& \int_{D_0} \{\fhat_0(s)-f_0(s)\} ds =n^{-1} \sum_{i=1}^n 2I(A_i=0)I(S_i \in D_0) -\int_{D_0}  f_0(s)ds\\
& := & n^{-1}\sum^{n}_{i=1}\phi_{K_1,i}+o_p\{n^{-1/2}\},\\
\hat{K}_2-K_2 &=& \int_{D_c} 2\frac{ f_0(s)}{f_1(s)}\{\fhat_0(s)-f_0(s)\} ds-\int_{D_c} \frac{f_0^2(s)}{f^2_1(s)} \{\fhat_1(s)-f_1(s)\} ds\\
&=&  n^{-1} \sum_{i=1}^n 2I(A_i=0)I(S_i \in D_c)2\frac{ f_0(S_i)}{f_1(S_i)} -\int_{D_c} 2\frac{ f^2_0(s)}{f_1(s)} ds\\
&&-\left\{n^{-1} \sum_{i=1}^n 2I(A_i=1) I(S_i \in D_c)\frac{f_0^2(S_i)}{f^2_1(S_i)} -\int_{D_c} \frac{f_0^2(s)}{f_1(s)}ds\right\} \\
& := & n^{-1}\sum^{n}_{i=1}\phi_{K_2,i}+o_p\{n^{-1/2}\}.
\end{eqnarray*}
Furthermore,
\begin{eqnarray*}
\lambdahat-\lambda&=& \frac{ \int_{D_c} \{\mhat_0(s)-\mhat_1(s)\}d{\Fhat}_{0}(s)}{\hat{K}_2+\hat{K}_1 \fhat_0(s^*)/\fhat_1(s^*)   } + \hat{K}_1 \frac{\mhat_0(s^*)\!-\!\mhat_1(s^*)}{\hat{K}_2+\hat{K}_1 \fhat_0(s^*)/\fhat_1(s^*)   } \\
&&- \frac{ \int_{D_c} \{m_0(s)-m_1(s)\}d{F}_{0}(s)}{K_2+K_1 f_0(s^*)/f_1(s^*) }- {K}_1 \frac{m_0(s^*)-m_1(s^*)}{{K}_2+{K}_1 f_0(s^*)/f_1(s^*)   }\\
&=& \frac{\left[\int_{D_c}\{\mhat_0(s)-m_0(s)-\mhat_1(s)+m_1(s)\}f_0(s)ds+\int_{D_c}\{m_0(s)-m_1(s)\}\{\fhat_0(s)-f_0(s)\}ds  \right]}{K_2+K_1 f_0(s^*)/f_1(s^*) } \\
&&+  {K}_1 \frac{\mhat_0(s^*)-m_0(s^*)-\mhat_1(s^*)+m_1(s^*)}{{K}_2+{K}_1 f_0(s^*)/f_1(s^*) } +\frac{m_0(s^*)\!-\!m_1(s^*)}{{K}_2+{K}_1 f_0(s^*)/f_1(s^*)  } n^{-1}\sum^{n}_{i=1}\phi_{K_1,i}  \\
&&- \frac{ \int_{D_c} \{m_0(s)-m_1(s)\}d{F}_{0}(s)+K_1 \{m_0(s^*)-m_1(s^*)\}}{\{K_2+K_1 f_0(s^*)/f_1(s^*)\}^2 }\\
&& \times  \left[\frac{1}{n}\sum^{n}_{i=1}\phi_{K_2,i}+\frac{1}{n}\sum^{n}_{i=1}\phi_{K_1,i}\frac{f_0(s^*)}{f_1(s^*)}+(nh)^{-1}\left\{ \frac{\sum_{i=1}^{n}\phi_{0,i} (s^*)}{f_1(s^*)}\!-\!\frac{f_0(s^*)\sum_{i=1}^{n}\phi_{1,i} (s^*)}{f_1^2(s^*)}\right\}\right]\\
&&+o_p\{(nh)^{-1/2}\}\\
&=& \frac{1}{K_2+K_1 f_0(s^*)/f_1(s^*)} \Bigg[ n^{-1}\sum^{n}_{i=1}2I(A_i=0)Y_i I(S_i \in D_c)-\int_{D_c}m_0(s)f_0(s)ds \\
&&-n^{-1}\sum^{n}_{i=1}2I(A_i=1)Y_i I(S_i \in D_c)f_0(S_i)/f_1(S_i) +\int_{D_c}m_1(s)f_0(s)ds\\
&&+n^{-1}\sum^{n}_{i=1}2I(A_i=0)I(S_i \in D_c)\{m_0(S_i)-m_1(S_i)\}-\int_{D_c}\{m_0(s)-m_1(s)\}f_0(s)ds \Bigg]\\
&&+ \frac{K_1}{{K}_2+{K}_1 f_0(s^*)/f_1(s^*) } (nh)^{-1}\sum^{n}_{i=1}\{\phi_{m,0,i}(s^*)-\phi_{m,1,i}(s^*)\} \\
&&+\frac{m_0(s^*)\!-\!m_1(s^*)}{{K}_2+{K}_1 f_0(s^*)/f_1(s^*)  } n^{-1}\sum^{n}_{i=1}\phi_{K_1,i} +o_p\{(nh)^{-1/2}\}\\
&&- \frac{ \int_{D_c} \{m_0(s)-m_1(s)\}d{F}_{0}(s)+K_1 \{m_0(s^*)-m_1(s^*)\}}{\{K_2+K_1 f_0(s^*)/f_1(s^*)\}^2 }\\
&& \times  \left[\frac{1}{n}\sum^{n}_{i=1}\phi_{K_2,i}+\frac{1}{n}\sum^{n}_{i=1}\phi_{K_1,i}\frac{f_0(s^*)}{f_1(s^*)}+(nh)^{-1}\left\{ \frac{\sum_{i=1}^{n}\phi_{0,i} (s^*)}{f_1(s^*)}\!-\!\frac{f_0(s^*)\sum_{i=1}^{n}\phi_{1,i} (s^*)}{f_1^2(s^*)}\right\}\right]\\
&:=&  (nh)^{-1}\sum^{n}_{i=1}\phi_{\lambda,i}+o_p\{(nh)^{-1/2}\}.
\end{eqnarray*}
Similarly, we can get
$\hat{c}-c=(nh)^{-1}\sum^{n}_{i=1}\phi_{c,i}+o_p\{(nh)^{-1/2}\}$.

Using above results we can obtain the influence functions for the optimal transformation function estimators by coupling delta method with the fact that
\begin{eqnarray*}
 \gopt(s)&=&\widetilde{G}\left(m_0(s), m_1(s), f_0(s), f_1(s), \lambda, c \right) \\
\text{and}\ \hat{g}(s)&=&\widetilde{G}\left(\mhat_0(s), \mhat_1(s), \fhat_0(s), \fhat_1(s), \lambdahat, \hat{c} \right).
\end{eqnarray*}
Specifically, we can show that
$$ \hat{g}(s)-\gopt(s)= (nh)^{-1}\sum_{i=1}^n \phi_{g,i}(s) +o_p\{(nh)^{-1/2}\},$$
where $E(\phi_{g, i}^2(s))<\infty$.

\clearpage

\bibliographystyle{biometri}
\bibliography{ref}

\end{document}